# Improved step-by-step chromaticity compensation method for chromatic sextupole optimization*

LIU Gang-Wen(刘刚文), BAI Zheng-He(白正贺)[1)], JIA Qi-Ka(贾启卡), LI Wei-Min(李为民), WANG Lin(王琳)

National Synchrotron Radiation Laboratory, University of Science and Technology of China, Hefei 230029, China

**Abstract:** The step-by-step chromaticity compensation method for chromatic sextupole optimization and dynamic aperture increase was proposed by E. Levichev and P. Piminov (E. Levichev and P. Piminov, 2006 [1]). Although this method can be used to enlarge the dynamic aperture of storage ring, it has some drawbacks. In this paper, we combined this method with evolutionary computation algorithms, and proposed an improved version of this method. In the improved method, the drawbacks are avoided, and thus better optimization results can be obtained.

## 1  Introduction

In storage rings for synchrotron radiation sources and high-energy physics, quadrupoles for beam focusing generate negative natural chromaticity that is required to be compensated for by chromatic sextupoles, which introduce nonlinearity into particle dynamics, limiting the region of stable motion of the particles, i.e., the so-called dynamic aperture (DA). To enlarge DA for increasing beam injection efficiency and beam lifetime, chromatic sextupoles need to be optimized. Techniques for chromatic sextupole optimization have been developed, including analytic and numerical ones. In the traditional analytic approach, one needs to minimize the strengths of resonances nearby, and control amplitude and momentum dependent tune shifts away from these resonances. This kind of optimization work is not direct, and the optimization results depend on one's experience. The popular and powerful numerical method developed in the recent years is to use evolutionary computation algorithms, such as genetic algorithm (GA) and particle swarm optimization (PSO). This method is direct and can quickly find global optimal solutions. Another numerical method that can also obtain global optimal solutions is the scanning method, which is very simple but only suitable for lattices with fewer families of chromatic sextupoles due to the problem of the huge amount of computation.

Besides, there is another simple numerical method for chromatic sextupole optimization but without having the computation problem, which was proposed by E. Levichev and P. Piminov [1] and slightly earlier than the application of GA to sextupole optimization and applied to the DA optimization for ALBA and DAΦNE upgrade [1-3]. In this method, chromaticity is corrected step-by-step along a straight line from the point of natural chromaticity to the point of desired chromaticity. At each step, only a small fraction of chromaticity is corrected with the best pair of focusing and defocusing chromatic sextupoles that makes DA largest at the step. This method can find larger DA, but it has some drawbacks. First, the optimization results depend on the number of steps. A larger number of steps can make the optimized DA better, but too large number would lead to reduction of the optimized DA as mentioned in Ref. [1]. Second, the off-momentum DAs can not be directly included into the optimization, this is because during step-by-step compensating for chromaticity the tunes of off-momentum particles are changing that may cause resonance crossing so that the off-momentum DA optimization is affected.

*Supported by NSFC 11175182 & 11175180.
1) E-mail: baizhe@ustc.edu.cn

In this paper, we considered the Levichev and Piminov method from a new point of view, and then pointed out that the drawbacks of this method can be avoided by introducing evolutionary computation algorithms, and thus proposed an improved version of this method. In this paper, PSO is adopted due to its faster convergence than GA (see for example Ref. [4]). A lattice of HLS-Ⅱ storage ring is taken as an example to illustrate the superiority of the improved method over the original one.

## 2 Improved method and application
### 2.1 Original method revisited

To better introduce our improved method, we first apply the Levichev and Piminov method to a non-achromatic lattice of HLS-Ⅱ storage ring from Ref. [5] with natural emittance of 14.9 nm-rad at 800 MeV. The lattice has two families of focusing chromatic sextupoles (S1 and S3) and two defocusing families (S2 and S4). Thus there are four pairs of focusing and defocusing chromatic sextupoles, i.e., (S1, S2), (S1, S4), (S3, S2) and (S3, S4). For this small ring lattice with horizontal and vertical natural chromaticities of (-14.44, -14.80), we found the best DA is obtained at the number of steps $N = 25$. Fig. 1 shows the optimized DA area values at $N = 8, 15, 20, 25$, and 75. We can see that larger DA can be obtained at suitably large N, but not at too large N (e.g., $N = 75$). This is because at too large N, the DA change from step to step is not obvious and the choice of the best pair becomes somewhat random as explained in Ref. [1].

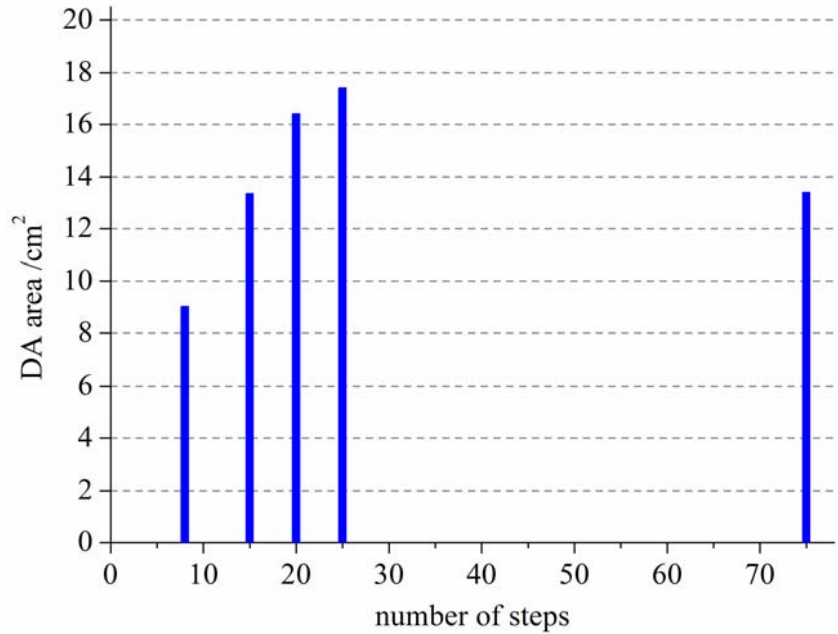

Fig. 1. The optimized DA areas at some numbers of steps.

Now consider only the best case of N = 25. After the optimization, the number of times that each pair of chromatic sextupoles is used as the best pair can be obtained as shown in Fig. 2. We can see that, for example, the pair (S1, S2) is used 16 times, which means that it contributes 16/25 of the total chromaticity compensation. Fig. 3 shows the best pair chosen at each step number during the optimization procedure. Imagine that if we change the order that these pairs are used in Fig. 3 while maintaining their numbers of times used in Fig. 2, the DA obtained must be the same. For example, at $1^{st}$ step we can use the pair (S1, S4) and meanwhile at $16^{th}$ step we use the pair (S1, S2). This is because the contribution of each pair to the total chromaticity compensation remains the same, or further, in other words, the strength of each sextuple family is unchanged under the change of the order. So we can care only about the number of times that each pair is used, but not the procedure how to obtain them. Further, we can use other procedures to replace the step-by-step procedure in the Levichev and Piminov method to obtain the number of times that each pair is used, for example as in Fig. 2.

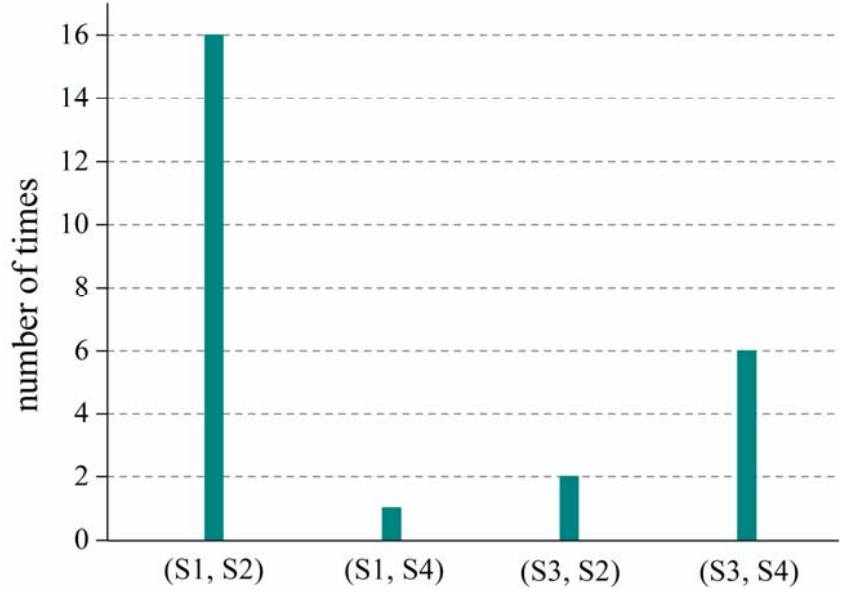

Fig. 2. The number of times that each pair is used in the best case of N = 25.

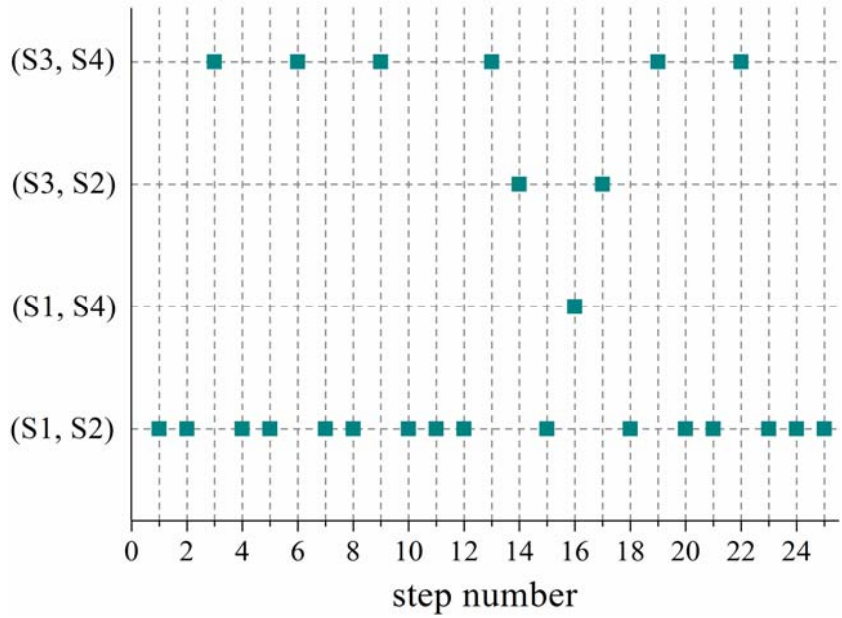

Fig. 3. The best pair chosen at each step number in the best case of N = 25.

### 2.2 Improved method

For a storage ring lattice with $N_F$ families of focusing chromatic sextupoles and $N_D$ families of defocusing chromatic sextupoles, there are $M = N_F \times N_D$ pairs of focusing and defocusing chromatic sextupoles. Let N be the number of steps adopted in the Levichev and Piminov method. After optimization using this method, the number of times that each pair is used as the best pair can be obtained, which we denote as $(N_1, N_2, \ldots, N_M)$. Of course, they satisfy the following equality:

$$N_1 + N_2 + \ldots + N_m + \ldots + N_M = N. \qquad (1)$$

If we do not care about how they are obtained, and imagine that the same $(N_1, N_2, \ldots, N_M)$ is obtained by another procedure, then the DA optimized by this procedure is the same as that optimized by the procedure in the Levichev and Piminov method. So from this point of view, ($N_1$, $N_2$, …, $N_M$) can be seen as optimization variables, and the procedure in the Levichev and Piminov method can be seen as an algorithm for optimizing ($N_1$, $N_2$, …, $N_M$) to enlarge DA, the optimization objective. Further, we can also use other algorithms to optimize the variables ($N_1$, $N_2$, …, $N_M$). Obviously, this kind of optimization problem can be well solved by the widely used evolutionary computation algorithms, such as GA and PSO.

Inspired by the idea above, we proposed an improved version of the Levichev and Piminov method, which is converted to a common optimization problem. In the improved method, there are M optimization variables, i.e., $N_1$, $N_2$, …, $N_M$, which are non-negative integers and subject to the equality constraint (1) for given N; and the optimization objective is DA, which is optimized using evolutionary computation algorithms. In other words, for a given N, any ($N_1$, $N_2$, …, $N_M$) satisfying the equality constraint (1) is a potential solution to the DA optimization problem, and the evolutionary computation algorithm is employed to search for the best solution ($N_1$, $N_2$, …, $N_M$) that has the largest DA. In this paper, the evolutionary computation algorithm we used is the PSO algorithm, which usually converges faster than GA. Taking the HLS-II storage ring lattice used above with four pairs of chromatic sextupoles as an example, and given N = 25, then for example (3, 9, 7, 6) is a potential solution to the DA optimization, which means that, for example, $1^{st}$ pair contributes 3/25 of the total chromaticity correction.

In the improved method, due to that larger N enhances the precision of the fraction $N_m/N$ that each pair contributes to the total chromaticity correction, we can set N very large to obtain better DA, but meanwhile without increasing the amount of computation. Besides, in the improved method, for each solution ($N_1$, $N_2$, …, $N_M$), when its DA is tracked the chromaticity has been corrected to the desired value with the solution. Thus the problem of resonance crossing for off-momentum particles is avoided, and the off-momentum DA or momentum aperture can be directly included into the optimization. In a word, the improved method we proposed overcomes the drawbacks that the original Levichev and Piminov method has, which will be demonstrated by applying the improved method to the lattice used above.

### 2.3  Application

Now we apply the improved method to the HLS-II storage ring lattice that we used for the original method. For this lattice with four pairs of focusing and defocusing chromatic sextupoles, there are four optimization variables, denoted as (N1, N2, N3, N4), which represent the contribution of each pair to the total chromaticity correction. We first optimize on-momentum DA, and the result is compared with that obtained using the original method. Then off-momentum DA is included, and on- and off-momentum DAs are simultaneously optimized. The PSO algorithm is employed for these optimizations.

First we consider only one optimization objective, the area of on-momentum DA. Three optimizations with N = 25, 75, 125 and 1,000,000 are done using the PSO algorithm with a population size of 20 that ran for 100 generations for each optimization to enlarge the DA area. The optimized DA area values are shown in Fig. 4. We can see that when N becomes larger, the optimized DA also becomes larger. This is due to the improvement of the precision of $N_m/N$. We can also see that the DA area value obtained with N = 125 is very close to that obtained with N = 1,000,000. This means that when N is larger than some value, the optimized DA area begins to converge.

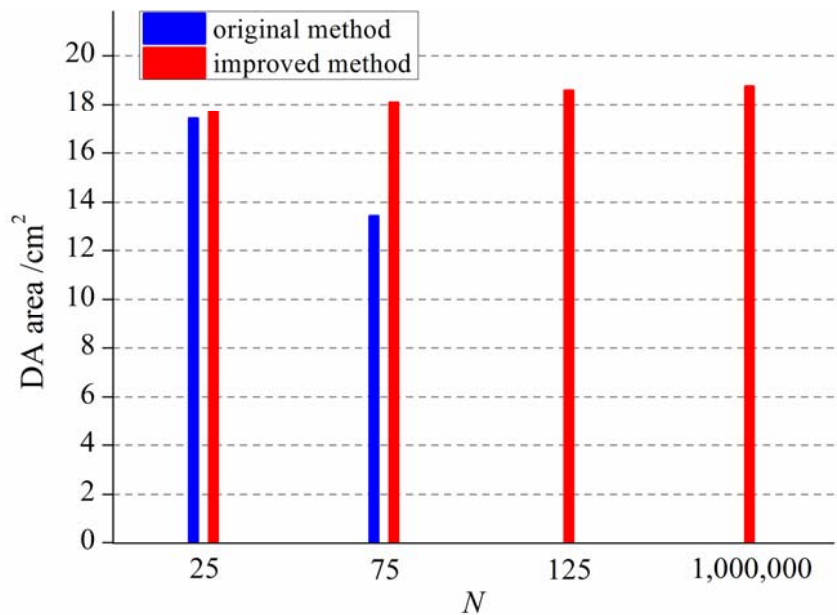

Fig. 4. The DA areas optimized using the improved method (red columns), compared with those obtained using the original method (blue columns). Note that the horizontal axis is not to scale.

Besides, as a comparison, the optimized DA areas obtained in Section 2.1 using the original method with N = 25 and 75 are also shown in Fig. 4. From Fig. 4, it can be seen that the DA area obtained using the improved method with N = 1,000,000 is larger than that obtained using the original method with the best N = 25, and these two DAs are shown in Fig. 5, from which it is clear to see that the former is larger than the latter. From Fig. 4, it can be also seen that even at N = 25, the DA obtained using the improved method is slightly larger than that obtained using the original method. If we also set N = 1,000,000 in the original method, not only better DA will not be obtained, but also the amount of computation is very huge. Fig. 6 shows the increase of the obtained best DA area with generation number for the optimization using the improved method with N = 1,000,000, from which we can see that the PSO algorithm begins to converge at about 20$^{th}$ generation. After the optimization, the best solution is obtained, which is shown in Fig. 7, representing the contribution of each pair to the total chromaticity correction.

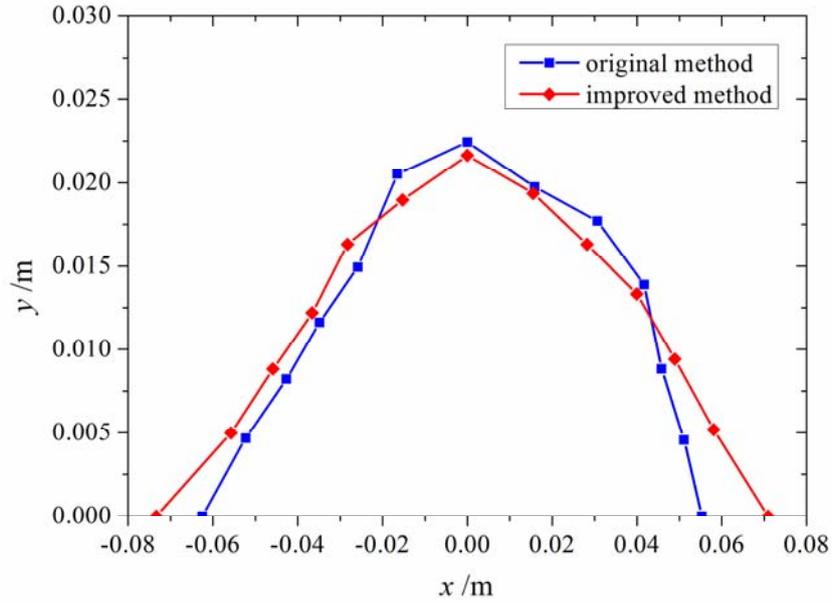

Fig. 5. The comparison of the optimized DAs obtained using the improved method with N = 1,000,000 (red line) and using the original method with the best N = 25 (bule line).

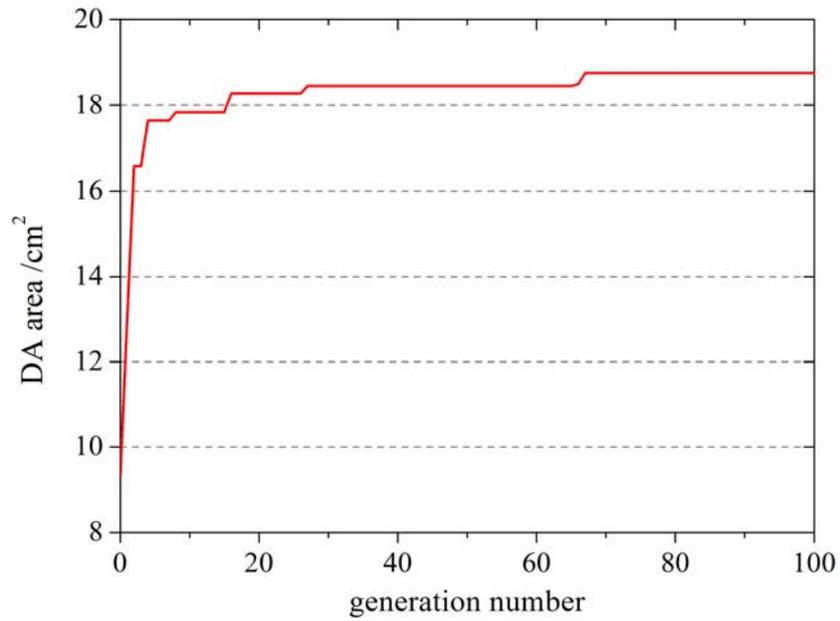

Fig. 6. The increase of the obtained best DA area with generation number for the optimization with N = 1,000,000.

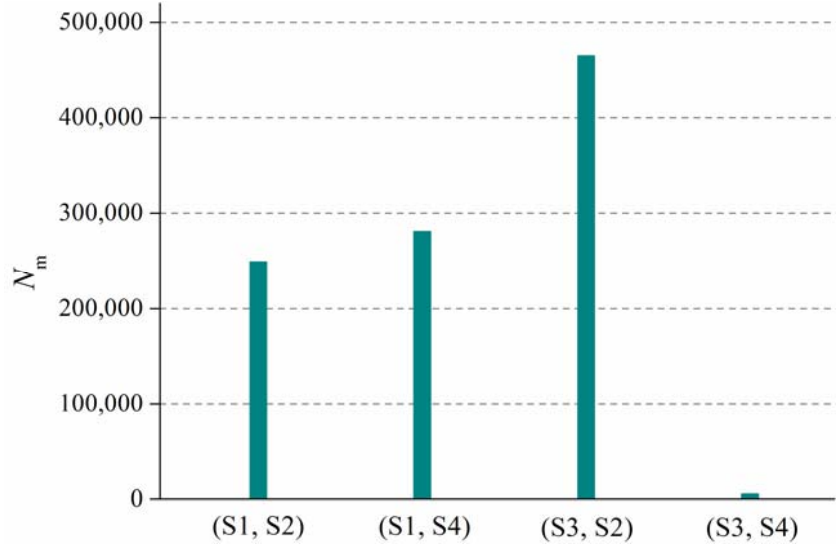

Fig. 7. The best solution obtained with N = 1,000,000.

Then we include off-momentum DA into the optimization for the lattice. The multi-objective PSO (MOPSO) [6] is employed to simultaneously optimize two optimization objectives, the on- and off-momentum (Δp/p = 2%) DA areas. The MOPSO algorithm with a population of 40 ran for 50 generations to enlarge the two DAs. The solutions together with Pareto optimal solutions obtained after the optimization are shown in Fig. 8, as well as the results obtained after initialization and $10^{th}$ generation. We can see that after the optimization, the DA areas of all solutions are on a much higher level than those in the initialization. One Pareto optimal solution is selected with both better on- and better off-momentum DAs, which is indicated by the black arrow in Fig. 8, and its DAs are shown in Fig. 9, from which we can see that the two DAs are all large. Fig. 10 shows the elements of this Pareto optimal solution, each of which representing the contribution of the corresponding pair to the total chromaticity compensation. In addition to off-momentum DA, of course, the momentum aperture (MA) can also be included into the optimization, and the DA and MA can also be simultaneously optimized using the improved method. But we do not do this work in this paper, since it is similar to the work we just presented.

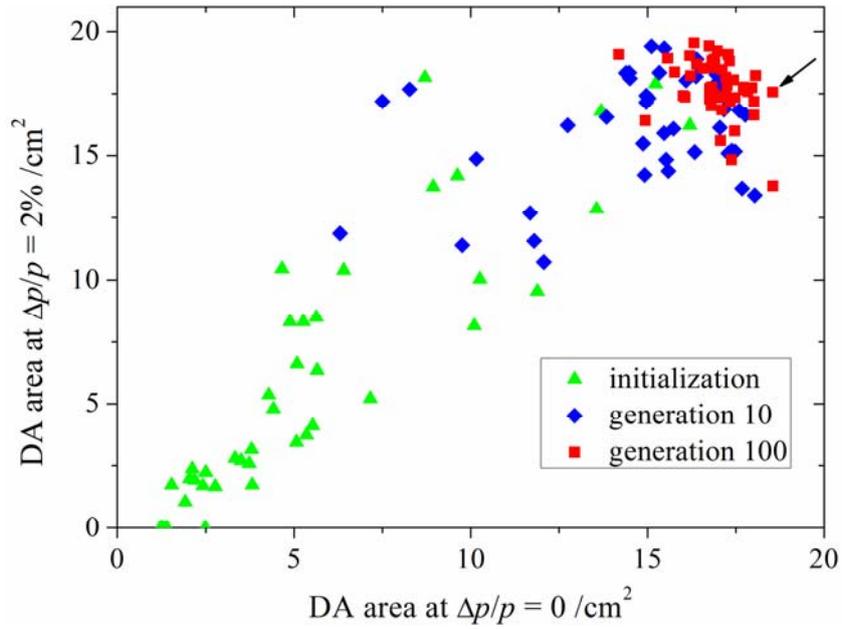

Fig. 8. The solutions together with Pareto optimal solutions obtained after initialization (green triangles), after 10[th] generation (blue diamonds) and after optimization (red squares).

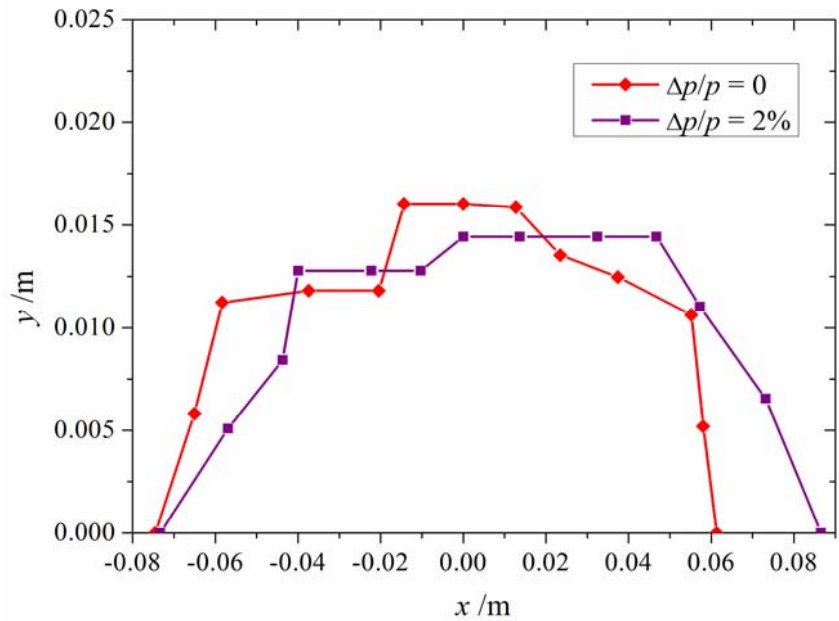

Fig. 9. The on- and off-momentum (2%) DAs of one Pareto optimal solution obtained with N = 1,000,000.

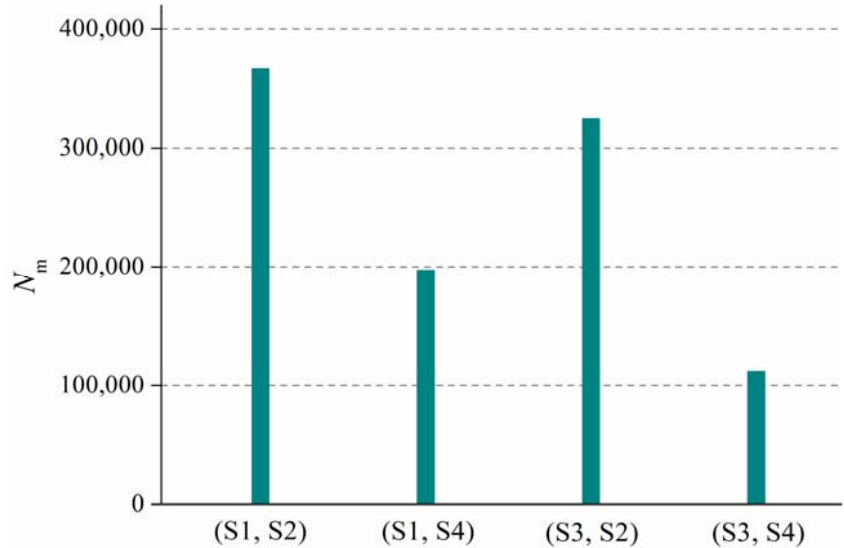

Fig. 10. The best solution obtained with N = 1,000,000 (the MOPSO case).

## 3 Conclusion

From a new point of view, we pointed out that the Levichev and Piminov method can be converted to a common optimization problem, where the optimization variables are the numbers related to the contribution of each pair of focusing and defocusing chromatic sextupoles to the total chromaticity compensation. Naturally, the evolutionary computation algorithms can be introduced to solve the optimization problem, and thus an improved Levichev and Piminov method was proposed, in which the drawbacks in the original Levichev and Piminov method can be avoided. In this paper, the PSO algorithm is adopted as the evolutionary computation algorithm, and a HLS-II storage ring lattice is used as an example of application of the improved method. The optimization results show that the improved method we proposed is better than the original method.